\documentclass[12pt]{iopart}
\usepackage{iopams}  
\usepackage{bm}
\usepackage{mathrsfs}
\usepackage[utf8]{inputenc}
\usepackage{enumerate}
\usepackage{appendix}
\usepackage{dcolumn}
\usepackage{multirow}
\usepackage{float}
\usepackage{color}
\begin{document}
\title[Towards a Gordon form of the Kerr spacetime]{
{Towards a Gordon form of the Kerr spacetime}}

\author{Stefano Liberati,$^{1,2}$ Giovanni Tricella,$^{1,2}$ and Matt Visser$^3$}

\address{$^1$ SISSA - International School for Advanced Studies, via Bonomea 265, \\
\qquad 34136 Trieste, Italy.}
\address{$^2$ INFN sezione di Trieste, via Valerio 2, Trieste, Italy.}
\address{$^3$ School of Mathematics and Statistics, Victoria University of Wellington, \\
\qquad PO Box 600, Wellington 6140, New Zealand.}
\ead{liberati@sissa.it, gtricell@sissa.it, matt.visser@sms.vuw.ac.nz}
\vspace{10pt}
\begin{indented}
\item[]11 March 2018; \LaTeX-ed \today
\end{indented}

\definecolor{purple}{rgb}{1,0,1}
\newcommand{\red}[1]{{\slshape\color{red} #1}}
\newcommand{\blue}[1]{{\slshape\color{blue} #1}}
\newcommand{\purple}[1]{{\slshape\color{purple} #1}}

\begin{abstract}
It is not currently known how to put the Kerr spacetime metric into the so-called Gordon form, although the closely related Kerr--Schild form of the Kerr metric is well known. 
A Gordon form for the Kerr geometry, if it could be found, would be particularly useful in developing analogue models for the Kerr spacetime, since the Gordon form is explicitly given in terms of the 4-velocity and ``refractive index'' of an effective medium. 
In the current article we report progress toward this goal. 
First we present the Gordon form for an approximation to Kerr spacetime in the slow-rotation limit, obtained by suitably modifying the well-known Lense--Thirring form of the slow-rotation metric. 
Second we present the Gordon form for the Kerr spacetime in the near-null limit, (the 4-velocity of the medium being close to null). 
That these two perturbative approximations to the Kerr spacetime in Gordon form exist gives us some confidence that ultimately one might be able to write the exact Kerr spacetime in this form.

coordinate transformation.  

\end{abstract}


\vspace{2pc}

\maketitle

\def\tr{{\mathrm{tr}}}
\def\cof{{\mathrm{cof}}}
\def\pdet{{\mathrm{pdet}}}
\def\d{{\mathrm{d}}}

%

\hrule
\tableofcontents
\markboth{Towards a Gordon form of the Kerr spacetime}
{Towards a Gordon form of the Kerr spacetime}
\bigskip
\hrule
\bigskip
\section{Introduction}

What is now called the ``Gordon form'' for a general class of spacetime metrics currently dates back some 95 years --- to 1923 --- originally being developed as a model for studying the propagation of light in a relativistic medium~\cite{Gordon}. 
A spacetime metric is said to be in Gordon form if:
\begin{equation}
g_{\mu\nu}=\eta_{\mu\nu}+(1-c_*^2) V_{\mu}V_{\nu}\, .
\label{eq:Gordon-Introduction}
\end{equation}
Here $\eta_{\mu\nu}$ is some background metric (typically taken to be flat Minkowski space), while
$V_\mu$ is some 4-velocity (properly normalized to $\eta^{\mu\nu} V_\mu V_\nu = -1$ in terms of the background metric), 
and  ${c_*}$ can be interpreted as the speed of light in the medium (so in terms of the refractive index $c_* = 1/n$). In situations discussed below (where we might not necessarily want to adopt the moving medium interpretation) $c_*$ can still be interpreted as the coordinate speed of light at spatial infinity.
This Gordon form for the spacetime metric has much deeper implications and a significantly wider range of applicability than the original context in which it was developed~\cite{Landau-Lifshitz,Plebanski:1960,Plebanski:1970,deFelice:1971,Skrotski,Balzas,Anderson,Pham,Schuster:2017,Schuster:2018,Vorticity,Boyer-Lindquist}, though  only relatively recently (2004) has it become clear that the theoretically important Schwarzschild spacetime can be put into this Gordon form~\cite{Rosquist,Liberati}. 

In counterpoint, the astrophysically and observationally more important Kerr spacetime (discovered in 1963~\cite{Kerr}) continues to attract considerable interest and provide unexpected new discoveries~\cite{Kerr,Kerr-book,Kerr-intro,Teukolsky:2014,Doran,River}.  One specific and relatively simple form of the Kerr metric is this
\begin{eqnarray}
\fl
\d s^2 &=&  -\mathrm{d}t^{2}+\mathrm{d}r^{2}+2a\sin^{2}\theta\,\mathrm{d}r\mathrm{d}\phi+\left(r^{2}+a^{2}\cos^{2}\theta\right)\,\mathrm{d}\theta^{2}+\left(r^{2}+a^{2}\right)\sin^{2}\theta\,\mathrm{d}\phi^{2}
\nonumber\\ \fl && \qquad
+\frac{2mr}{r^{2}+a^{2}\cos^{2}\theta} \left( -\mathrm{d}t\mp\mathrm{d}r\mp a\,\sin^{2}\theta\,\mathrm{d}\phi \right)^2.
\end{eqnarray}
Here $a=J/m$ is as usual the angular momentum per unit mass of the black hole.
Several other forms for the Kerr metric are known~\cite{Kerr-book,Kerr-intro,Teukolsky:2014}.
Indeed, only relatively recently (2000) was the Doran form of the Kerr spacetime developed~\cite{Doran,River}; this seeming to be as close as one can get to putting the Kerr metric into Painleve--Gullstrand ``acoustic geometry'' form. Now, historically it has been found that every significantly new form of the Kerr solution has lead to advances in our understanding, and it is still possible (though maybe not entirely likely) that the Kerr solution could be greatly simplified by writing it in some particularly clean form~\cite{Kerr-intro}. 

This raises the natural question --- is it possible to put the Kerr spacetime into Gordon form? One reason for being particularly interested in this question is the observation that many analogue spacetimes are quite naturally presented in Gordon form. (See particularly~\cite{Vorticity, Visser:2010}, or more generally~\cite{Unruh:1980,Visser:1993,Visser:1997} and~\cite{Barcelo:2000,Barcelo:2003,LRR,Visser:2001,Visser:2007,Visser:2013}.)
We shall partially answer this question by presenting two perturbative calculations. First we shall perform a slow-rotation calculation putting the well-known Lense--Thirring metric into Gordon form. 
Second we shall present a ``near-null'' version of Kerr in Gordon form; by starting with Kerr in Kerr--Schild form and performing an appropriate infinitesimal coordinate transformation.  In view of the black hole uniqueness theorems we know that the Kerr family is unique; we are not looking for a new spacetime. Instead we are looking for a new way of writing the quite standard Kerr spacetime; we are trying to find a coordinate transformation to simplify the presentation of the Kerr spacetime.

\section{Two easy results}
To set the stage, let us first present two simple results, before developing a general algorithm for implementing infinitesimal coordinate changes. 
\subsection{Gordon form of Schwarzschild spacetime}
The Gordon form of the Schwarzschild metric~\cite{Rosquist,Liberati} is less well-known than perhaps is should be.
Consider the line element
\begin{equation}
\fl
\d s^2 = \left(\eta_{\mu \nu} + \left(1-c_*^{2}\right) V_{\mu} V_{\nu}\right)\mathrm{d}x^{\mu}\mathrm{d}x^{\nu}; \qquad   
V = - \sqrt{1+\frac{2\tilde m}{r}}\,\mathrm{d}t+ \sqrt{\frac{2\tilde m}{r}}\, \mathrm{d}r.
\end{equation}
In spherical coordinates this is
\begin{equation}
\fl
\d s^2 = -\mathrm{d}t^2+\mathrm{d}r^2 + r^2(\mathrm{d}\theta^2+\sin^2\theta \mathrm{d}\phi^2) 
+ \left(1-c_*^2\right) \left( - \sqrt{1+\frac{2\tilde m}{r}}\,\mathrm{d}t+ \sqrt{\frac{2\tilde m}{r}}\, \mathrm{d}r\right)^2.
\label{eq:Gordon-Sch}
\end{equation}
This is spherically symmetric and easily checked to be Ricci flat --- so by Birkhoff's theorem it must be Schwarzschild spacetime in disguise. 
Here $c_*$ is an arbitrary constant $c_*\in(0,1)$, which at spatial infinity can be viewed, as anticipated, as the coordinate speed of light. Furthermore
 $V_a$ is a 4-velocity, (normalized in the background metric,  $\eta_{\mu\nu} V^\mu V^\nu = -1$), and the parameter $\tilde m$ is 
 proportional to the physical mass of the Schwarzschild spacetime.
By noting that
\begin{equation}
g_{tt} = -1 + (1-c_*^2)(1+2\tilde m/r) = -c_*^2 + (1-c_*^2)2\tilde m/r,
\end{equation}
and comparing to the asymptotic behaviour of Schwarzschild in the usual curvature coordinates,
we identify the physical mass as
\begin{equation}
m = {(1-c_*^2)\,\tilde m\over c_*^2} = (c_*^{-2}-1)\, \tilde m.
\end{equation}

\subsection{Gordon form of Lense--Thirring slow-rotation spacetime}

Let us remind ourselves of the quite standard version of the Lense--Thirring slow-rotation spacetime (in the usual Schwarzschild curvature coordinates). The line element is:
\begin{equation}
\fl\quad
\mathrm{d}s^2 = -\left(1-\frac{2m}{r}\right)\mathrm{d}t^2 + \left(1-\frac{2m}{r}\right)^{-1} \mathrm{d}r^2+ r^2 \mathrm{d}\theta^2 
+ r^2 \sin^2 \theta \left(\mathrm{d}\phi - {2ma\over r^3} \mathrm{d}t\right)^2.
\end{equation}
This represents a metric which is Schwarzschild (in curvature coordinates) plus $O(a)$ modifications, and for this metric one can easily check that $R_{ab} = O(a^2)$; all components of the Ricci tensor are $O(a^2)$. 
This $O(a^2)$ behaviour for the Ricci tensor is \emph{what we mean}  by saying that the Lense--Thirring spacetime is an approximate solution to the vacuum Einstein equations corresponding to a slowly rotating spacetime. 
The spacetime has angular momentum $J=ma$. 
For current purposes we could equally well ignore the $O(a^2)$ term \emph{in the metric} and write the simplified Lense--Thirring line element as:
\begin{equation}
\fl
\mathrm{d}s^2 = -\left(1-\frac{2m}{r}\right)\mathrm{d}t^2 + \left(1-\frac{2m}{r}\right)^{-1} \mathrm{d}r^2+ r^2 \mathrm{d}\theta^2 
+ r^2 \sin^2 \theta \mathrm{d}\phi^2   - {4ma\sin^2\theta\over r} \mathrm{d}t \mathrm{d}\phi \, . \label{Lense--Thirring}
\end{equation}
This simplified line element represents a metric which is still Schwarzschild (in curvature coordinates) plus $O(a)$ modifications, and for this metric we still get $R_{ab} = O(a^2)$; all components of the Ricci tensor are $O(a^2)$. That is, the spacetime is still Ricci flat up to terms quadratic in $a$. 

Based on these observations, to find a Gordon form for Lense--Thirring we simply take the Gordon form of Schwarzschild and make the ansatz
\begin{equation}
V = V_\mu \, \mathrm{d}x^\mu \; \to \; 
- \sqrt{1+{2\tilde m\over r} } \,\mathrm{d}t 
+\sqrt{2\tilde m\over r} \, \mathrm{d}r 
+ {2\tilde m\tilde{a}\sin^2\theta\over r\sqrt{1+\frac{2\tilde m}{r}}} \mathrm{d}\phi \, . \label{eq:Gordon Kerr small a 1-form}
\end{equation}
That is, we consider the metric ansatz represented by the line element
\begin{eqnarray}
\d s^2 &=& -\mathrm{d}t^2+\mathrm{d}r^2 + r^2(\mathrm{d}\theta^2+\sin^2\theta \mathrm{d}\phi^2) 
\nonumber\\
&&+ \left( 1-c_*^2 \right) \left(-\sqrt{1+{2\tilde m\over r} } \,\mathrm{d}t 
 +\sqrt{\frac{2\tilde m}{r}} \mathrm{d}r + {2\tilde m\tilde{a}\sin^2\theta\over r\sqrt{1+\frac{2\tilde m}{r}}} \mathrm{d}\phi\right)^2. \label{ansatz Gordon Lense--Thirring}
\end{eqnarray}
Here we have again $c_*\in(0,1)$, the parameter $\tilde m$ is proportional to the physical mass of the Lense--Thirring spacetime, and $\tilde{a}$ is proportional to $a$.
Note that (in the background metric) $||V||^2=-1+O(\tilde a^2)$, so that $V$  is approximately a unit timelike 4-vector.
Furthermore, since obviously $\tilde{a}=O(a)$, to first order in $a$ this metric ansatz is the just Gordon form of Schwarzschild plus an $O(a)$ perturbation. Finally,  a brief computation verifies that $R_{ab}=O(a^2)$, the metric is Ricci-flat to $O(a^2)$.  This observation justifies calling this metric the Gordon form of Lense--Thirring spacetime. 
That is, for slow rotation, we can approximate the Kerr spacetime to arbitrary accuracy by a metric that is of the Gordon form. 

To see how the parameters $\tilde a$ and $\tilde m$ are related to the physical parameters $a$ and $m$, note that at very large $r$ we have  $g_{tt}\to -c_*^2$, while at all values of $r$ we have $g_{t\phi} =  2(1-c_*^2)\tilde m\tilde a \sin^2\theta/r$.  Comparing this to the equivalent results for the usual form of the Lense--Thirring line element, (where at very large $r$ we have  $g_{tt}\to -1$, while at all values of $r$ we have $g_{t\phi} =  2ma \sin^2\theta/r$), we see that:
\begin{equation}
J = ma = {(1-c_*^2)\tilde m\tilde a\over c_*} = {(1-c_*^2)\tilde m\over c_*^2} \times (c_* \,\tilde a ) 
= m \times (c_* \, \tilde a).
\end{equation}
That is, $a = c_*\, \tilde a$, while $m=(c_*^{-2}-1)\tilde m$.

\section{General algorithm} 
Now let us try to make these observations more systematic by presenting a general algorithm for searching for the Gordon form (if it exists).

\subsection{Non-normalized Gordon and Kerr--Schild forms}

Both Gordon and Kerr--Schild forms of the metric express the metric tensor as the sum of a Riemann-flat
background metric $\overline{g}_{\mu\nu}$ and a $1$-form $v_\mu$ in tensor product with itself. Let us
adopt the notation
\begin{equation}
g_{\mu\nu}=\overline{g}_{\mu\nu}+v_{\mu}v_{\nu}\label{eq:Gordon}\,.
\end{equation}
Here $v_\mu$ is not normalized; this lack of normalization is useful in some explicit computations.
If $v$ is timelike (with respect to the background metric) then the Gordon form expression of equation~\eref{eq:Gordon-Introduction} can be recovered  normalizing $v_\mu = ||v|| V_\mu$.
If $v$ is null then we call this a Kerr--Schild form for the metric tensor. (The remaining case where $v$ is spacelike does not seem to be particularly interesting.)
In general, letting $\overline{g}^{\mu\nu}$ denote the inverse of the flat background
metric, which here we do not necessarily presume has to be in the form $\eta_{\mu\nu}={\rm diag}(-1,+1,+1,+1)$, 
the inverse of the full metric is
\begin{equation}
g^{\mu\nu}=\overline{g}^{\mu\nu}
-\frac{\overline{g}^{\mu\alpha}\;\overline{g}^{\nu\beta}\;v_{\alpha}\, v_{\beta}}
{1+\overline{g}^{\rho\sigma}\;v_{\rho}\,v_{\sigma}}\, .
\label{eq:Inverse metric Gordon}
\end{equation}
The specific choice of coordinates is manifestly irrelevant for this description: As long as the metric tensor can be put into
a Gordon form, every coordinate transformation \emph{that acts on both sides} will provide
an equivalent expression for the same decomposition.
It is in principle possible to find inequivalent Gordon forms for the same spacetime
if, choosing a common flat background metric, different $1$-forms
provide different full metric tensors which are equivalent through
coordinate transformations.

\subsection{How to find analytic expressions for Gordon and Kerr--Schild forms}
\label{subsec:How-to-find}

Knowing an expression for the full metric in a certain set of coordinates
$g_{\mu\nu}$, and an expression for the flat metric in a generally
different set of coordinates $\overline{g}_{\alpha\beta}$, we look
for possible inequivalent Gordon forms of the metric by applying a coordinate transformation
of the form
\begin{eqnarray}
\qquad \qquad \, \, x^{\mu}\rightarrow x^{\prime\mu} &=  x^{\mu}+\xi^{\mu}\left(x\right)\, ,
\label{eq:Coordinate local translation coordinate transformation}
\end{eqnarray}
and then noting
\begin{eqnarray}
g_{\mu\nu}\!\left(x\right)\mathrm{d}x^{\mu}\mathrm{d}x^{\nu}&\rightarrow& g_{\mu\nu}\left(x^{\prime}\right)\mathrm{d}x^{\prime\mu}\mathrm{d}x^{\prime\nu}
\nonumber\\
&=&  g_{\mu\nu}\left(x+\xi\right)\left(\delta_{\alpha}^{\mu}+\frac{\partial\xi^{\mu}}{\partial x^{\alpha}}\right)\left(\delta_{\beta}^{\nu}+\frac{\partial\xi^{\nu}}{\partial x^{\beta}}\right)\mathrm{d}x^{\alpha}\mathrm{d}x^{\beta}\, .
\label{eq:Metric tensor local translation coordinate transformation}
\end{eqnarray}
The RHS of equation~\eref{eq:Metric tensor local translation coordinate transformation}
is a new expression for the spacetime metric which depends
on the chosen local translations $\xi$ defining the coordinate transformations of
equation~\eref{eq:Coordinate local translation coordinate transformation}. This expression
can be written in a Gordon form if it is possible to find a $1$-form
$v$ satisfying
\begin{eqnarray}
g_{\mu\nu}\left(x+\xi\right)\left(\delta_{\alpha}^{\mu}+\partial_{\alpha}\xi^{\mu}\right)\left(\delta_{\beta}^{\nu}+\partial_{\beta}\xi^{\nu}\right)-\overline{g}_{\alpha\beta}\left(x\right) & =  v_{\alpha}v_{\beta}\, .
\label{eq:System of equations for Gordon form}
\end{eqnarray}
It is a straightforward algebraic exercise to extract --- up to an overall sign --- the expressions for the functions
$v_{\alpha}$ in terms of the functions $\xi^{\mu}$ and their derivatives,
from four of these ten equations. The remaining six equations provide a system
of highly non-trivial and non-linear partial differential equations for the functions
$\xi^{\mu}$ and the initially chosen tensors $g$ and $\overline{g}$.

Ultimately the problem of finding a Gordon form for the metric is
 that of finding an appropriate coordinate transformation,
\emph{i.e.}, solving the differential equations for the $\xi^{\mu}$, such
that the initial system equation~\eref{eq:System of equations for Gordon form}
admits a solution.
In particular, it will be operationally convenient to investigate a class
of coordinate transformations which is general enough to find a solution, but
possibly without spoiling the explicit symmetries of the metric. 
The Schwarzschild spacetime is a remarkable example of a system where
this problem admits an explicit solution, and we first use this to present a specific implementation of the general algorithm.

\subsection{Checking Schwarzschild in Kerr--Schild form}

We first apply this procedure to recover the well known Kerr--Schild form of the Schwarzschild metric,
describing a static black hole of physical mass $m$. The usual expression for the metric
obviously requires a transformation of coordinates to be put in a
Gordon form since when choosing spherical coordinates for the flat background (mildly abusing notation by conflating metrics with their line elements)
\begin{eqnarray}
\overline{g}_{\rm spherical} & = -\mathrm{d}t^{2}+\mathrm{d}r^{2}+r^{2}\mathrm{d}\theta^{2}+r^{2}\sin^{2}\theta\mathrm{d}\phi^{2}\, ,
\label{eq:Spherical flat metric-1}
\end{eqnarray}
we have 
\begin{eqnarray}
g & = -\left(1-\frac{2m}{r}\right)\mathrm{d}t^2 + \left(1-\frac{2m}{r}\right)^{-1} \mathrm{d}r^2+r^{2}\mathrm{d}\theta^{2}+r^{2}\sin^{2}\theta\mathrm{d}\phi^{2}
\label{eq:Schwarzschild standard metric}\\
 & \neq  \overline{g}_{\rm spherical}+v\otimes v \, .
 \label{eq:Generic Gordon form spherical background}
\end{eqnarray}
We need to apply a coordinate transformation which does not spoil the explicit
time translation symmetry and the explicit spherical symmetry. The
spatial coordinates are better left untouched since the angular part
of both the full metric and the spherical flat background is the same:
$v$ must have vanishing angular components, making a rotation completely
irrelevant; moreover, $r^{2}$ appears as the scale factor of the
angular part of both metric tensors, and therefore $r$ cannot be
transformed.
So we shall initially consider the simple coordinate transformation $t\rightarrow t+f\left(r\right)$. 

Applying such transformation to the Schwarzschild metric, and choosing as 
flat background the spherical flat metric $\overline{g}_{\rm{spherical}}$,  
the system in equation~\eref{eq:System of equations for Gordon form} admits a 
solution for $f^{\prime}\left( r \right)= \mp \frac{2m}{r-2m}$. That is,
we obtain the Kerr--Schild expression for the Schwarzschild spacetime,
as equation~\eref{eq:Gordon}, with
\begin{eqnarray}
g & = \overline{g}_{\rm spherical}+v\otimes v\, ,
\label{eq:Kerr-Schild Schwarzschild}\\
v & = -\sqrt{\frac{2m}{r}}\,\mathrm{d}t\mp\sqrt{\frac{2m}{r}}\,\mathrm{d}r\, .
\label{eq:Kerr-Schild Schwarzschild 1-form}
\end{eqnarray}
The $1$-form $v$ defines a Kerr--Schild decomposition since it is a null 1-form, 
as is easily checked by verifying $\overline{g}^{\mu\nu}v_{\mu}v_{\nu}=0$. The overall sign is 
chosen in such a way that the dual of this $1$-form is a future-directed vector field.

In this case the process of solving the system of equations~\eref{eq:System of equations for Gordon form}  
only requires the expression for $f'(r)$, which can be obtained algebraically: The 
analytical expression for $f(r)$ itself is not needed; it is enough to know that $f'(r)$ 
is integrable to be sure that the coordinate transformation is properly defined.
The computation was made particularly easy by an appropriate
choice of background: The flat metric in spherical coordinates explicitly
contains the same symmetries as the full spacetime, and this background
is simply the limit of the full initial metric for vanishing black
hole mass $m\rightarrow0$.
(This can also be considered as the limiting case of vanishing angular momentum for the known solution of the Kerr--Schild form of the Kerr metric.)

\subsection{Checking Schwarzschild in Gordon form}

A second application of the general algorithm allows us to recover the known result of the Gordon form 
of the Schwarzschild metric: by considering a somewhat more general class of coordinate transformations it is
possible to find inequivalent non-null $1$-forms reproducing the
Schwarzschild metric as in the Gordon form of equation~\eref{eq:Gordon}.

The reasoning  presented above suggested that it would be profitable to consider a translation of the $t$ coordinate by a function of the 
radial coordinate $r$ only, so that the explicit symmetries of the metric were preserved. However, more generally we note that the time translation symmetry 
is still explicitly preserved if the $t$ coordinate is deformed by rescaling. So a wider class of coordinate transformations
to apply to the standard Schwarzschild metric equation~\eref{eq:Schwarzschild standard metric} is this
\begin{equation}
 t\to\sqrt{1-\zeta}\,t+f\left(r\right)\, ;\\
 \mathrm{d}t\to\sqrt{1-\zeta}\, \mathrm{d}t+f^{\prime}\left(r\right)\mathrm{d}r\, .
 \end{equation}
 We will consider the rescaling factor $\sqrt{1-\zeta}=c_*$ many times in the following discussion, and we will usually refer to $\zeta$ as the deformation parameter. 
This class of coordinate transformations modifies the appearance of the metric tensor, and the metric can be written in Gordon form with respect to the flat spherical background, 
(that is, a solution for the system of equations~\eref{eq:System of equations for Gordon form} with $\overline{g}=\overline{g}_{\rm spherical}$ exists), if and only if
\begin{eqnarray}
f^{\prime}\left(r\right) & =  \mp\frac{2m}{r-2m}\sqrt{1+\zeta \frac{r-2m}{2m}}\, .
\label{eq:Gordon Schwarzschild coordinate transformation}
\end{eqnarray}
Since $f'(r)$ is integrable this describes a proper coordinate transformation.\\

In conclusion, the Schwarzschild metric can be cast in a Gordon form with
\begin{eqnarray}
g & =  \overline{g}_{\rm spherical}+v\otimes v\, ,\label{eq:Gordon Schwarzschild}\\[5pt]
 v&=-\sqrt{\zeta+\frac{2m}{r}\left(1-\zeta\right)}\;\mathrm{d}t
 \mp\sqrt{\frac{2m}{r}\left(1-\zeta\right)}\;\mathrm{d}r \, .\label{eq:Gordon Schwarzschild 1-form}
\end{eqnarray}
The $1$-form $v$ is in general non-null, since $\overline{g}^{\mu\nu}v_{\mu}v_{\nu}=-\zeta$;
the limit $\zeta\rightarrow0$ reproduces the Kerr--Schild form.
The original expression of the Gordon form  of equation~\eref{eq:Gordon-Introduction}, or equation~\eref{eq:Gordon-Sch}, is obtained by rewriting these expressions in terms of  the speed of light in the medium and the normalized 4-velocity:
\begin{eqnarray}
c_*^2 & = 1-\zeta \, , \qquad  \zeta = 1-c_*^2,
\\
V & =\frac{v}{\sqrt{\zeta}}= -\sqrt{1+\frac{2\tilde m}{r}}\;\mathrm{d}t \mp \sqrt{\frac{2\tilde m}{r}}\;\mathrm{d}r\,,
\\
\tilde m & = m \;\frac{1-\zeta}{\zeta}\,.
\end{eqnarray}
Here $m$ is again the physical mass, while $\tilde m$ is a convenient shorthand.
The parameter $\zeta$ is bounded from both sides: In order to transform
the $t$ coordinate we must have $\zeta<1$, otherwise we wouldn't
be able to consider the square root $\sqrt{1-\zeta}$. From equation~\eref{eq:Gordon Schwarzschild coordinate transformation}
we also understand that it must be required that the parameter $\zeta$ be
non-negative: In order for the square root $\sqrt{1+\zeta \frac{r-2m}{2m} }$
to exist in the external region $r>2m$, we must have $\zeta\geq0$.
In conclusion, the deformation parameter is bounded within a finite
interval $\zeta\in\left[0,1\right)$, which corresponds to $c_*^2\in(0,1]$, the speed of light in the medium should be real, non-negative, and bounded by the speed of light in vacuum.  

\section{The Kerr spacetime
\label{sec:The-Kerr-spacetime}}
We now consider the more interesting case of rotating black holes, described
with the Kerr metric, which we would like to express in Gordon form.

It is well-known that the Kerr metric can be written in Kerr--Schild
form, which we could obtain following an analogous procedure to the
one seen for the Schwarzschild spacetime. For our purposes, it is
most convenient to consider the expression of the metric tensor as
presented in Kerr's original derivation~\cite{Kerr}. Again slightly abusing notation by conflating the metric with its corresponding line element we have
\begin{eqnarray}
\fl g =&\left(r^{2}+a^{2}\cos^{2}\theta\right)\left(\mathrm{d}\theta^{2}+\sin^{2}\theta\mathrm{d}\phi^{2}\right)+2\left(\mathrm{d}u+a\sin^{2}\theta\mathrm{d}\phi\right)\left(\mathrm{d}r+a\sin^{2}\theta\mathrm{d}\phi\right)
\nonumber \\
\fl &-\left(1-\frac{2mr}{r^{2}+a^{2}\cos^{2}\theta}\right)\left(\mathrm{d}u+a\sin^{2}\theta\mathrm{d}\phi\right)^{2} \, ,
\end{eqnarray}
where $u$ should be read as a modified time coordinate (the advanced/retarded time). Applying the transformation $u\rightarrow\pm t+r$ to this metric
we easily obtain  the Kerr--Schild form of the Kerr metric,
making use of a non-trivial representation of the flat background
\begin{eqnarray}
g & =  \overline{g}_{\rm spheroidal}+v\otimes v \, .
\label{eq:Kerr-Schild Kerr}
\end{eqnarray}
Here
\begin{eqnarray}
\fl
\overline{g}_{\rm spheroidal} & =  -\mathrm{d}t^{2}+\mathrm{d}r^{2}+2a\sin^{2}\theta\,\mathrm{d}r\mathrm{d}\phi+\left(r^{2}+a^{2}\cos^{2}\theta\right)\,\mathrm{d}\theta^{2}+\left(r^{2}+a^{2}\right)\sin^{2}\theta\,\mathrm{d}\phi^{2}, 
\nonumber\\ \fl &\quad
\label{eq:Kerr-Schild Kerr background}
\end{eqnarray}
is a  non-trivial non-diagonal implementation of spheroidal coordinates, and the null 1-form $v$ is given by
\begin{eqnarray}
v & =  \sqrt{\frac{2mr}{r^{2}+a^{2}\cos^{2}\theta}}\,\left( -\mathrm{d}t\mp\mathrm{d}r\mp a\,\sin^{2}\theta\,\mathrm{d}\phi \right)\, .\label{eq:Kerr-Schild Kerr 1-form}
\end{eqnarray}

\medskip
As expected, the limit $a\rightarrow0$ of $\overline{g}_{\rm spheroidal}$ yields the spherical polar flat metric,
while the limit $a\rightarrow0$ of  $v$ reproduces the $1$-form
of equation~\eref{eq:Kerr-Schild Schwarzschild 1-form} for the Kerr--Schild
form of the Schwarzschild metric, as presented in equation~\eref{eq:Kerr-Schild Schwarzschild}.
This decomposition is therefore a general description for Kerr,
of which the Schwarzschild version is a particular case. The norm of $v$ can be 
shown --- after computing $\overline{g}_{\rm spheroidal}^{\mu\nu}$ ---
to be vanishing, proving that this expression is indeed of Kerr--Schild form.

The flat background metric~\eref{eq:Kerr-Schild Kerr background}
considered here is the limit $m\rightarrow0$ of the full metric; it is
indeed Riemann-flat since it is obtained from the usual spherical
flat metric equation~\eref{eq:Spherical flat metric-1} through a (somewhat non-obvious) coordinate transformation
\begin{eqnarray}
\qquad r^{2} & \rightarrow & r^{2}+a^{2}\sin^{2}\theta\, ,\\
\quad \sin^{2}\theta & \rightarrow & \frac{\left(r^{2}+a^{2}\right)\sin^{2}\theta}{r^{2}+a^{2}\sin^{2}\theta} \, ,\\
\qquad\phi & \rightarrow & \phi+\arctan\left(\frac{r}{a}\right)\, .\label{eq: phi transformation}
\end{eqnarray}
Note that not transforming the coordinate $\phi$ as done in equation~\eref{eq: phi transformation}
would have resulted in obtaining the somewhat more usual diagonal form of oblate spheroidal coordinates (without the $\d r\, \d\phi$ cross term).

To apply the general procedure described in subsection \ref{subsec:How-to-find} and to search for 
a Gordon form of the Kerr metric, we now need to manipulate the metric tensor with a sufficiently wide class
of coordinate transformations. 
Indeed, this will be considerably less trivial than for the Schwarzschild metric, simply because the Kerr spacetime has fewer symmetries. 
This implies that, while previously we could assume transformations preserving spherical symmetry, now in 
the Kerr spacetime we can only make weaker assumptions, as only an axial symmetry
is left in the spatial sector. 
Expressing the coordinate transformation in terms of local
translations as in equation~\eref{eq:Coordinate local translation coordinate transformation},
we can expect the translation to depend neither on $\phi$, nor on $t$,
apart from again possibly rescaling $t$  by the factor $\sqrt{1-\zeta}=c_*$. Accordingly we consider
\begin{equation}
x^{\mu}\rightarrow x^{\mu}+\left(\sqrt{1-\zeta}-1\right)\delta_t{}^{\mu} \; t+G^{\mu}\left(r,\theta\right)
\, .
\end{equation}
In general all four coordinates should now be transformed.

As discussed in subsection \ref{subsec:How-to-find}, the full
resolution of the system of equations~\eref{eq:System of equations for Gordon form}
requires solving nonlinear (quadratic) partial differential equations. Therefore
finding the exact Gordon form for the Kerr metric seems (at least for now) to be a step too far. 
But we can certainly investigate this system perturbatively: We can expect that
for a small rotation parameter $a$ the perturbative expression of the Gordon
form of the Kerr spacetime is a perturbation of the Gordon form of
the Schwarzschild spacetime.  Moreover, for a small deformation of
the time coordinate --- thereby considering a description
at first order in $\zeta$ --- we can also expect the Gordon
form of the Kerr spacetime to be a perturbation of the known Kerr--Schild
form of the Kerr metric.

In the following analysis we present these two different perturbative approaches,
and the general expressions resulting from the resolution of the system~\eref{eq:System of equations for Gordon form}
within the two separate approximations of slow-rotation and a near-null 1-form. 
\subsection{Slow rotation}
When considering the slow-rotation regime, we can adopt approximate descriptions of the Kerr metric
equivalent to the Lense--Thirring metric~\eref{Lense--Thirring},  which can be interpreted 
as a perturbation of the Schwarzschild metric.
So in this subsection we shall consider a perturbative expansion for small angular momentum: We look for a Gordon form approximating
the Kerr metric at first order in $a$. This is a common assumption in the 
literature, which is physically reasonable since astrophysical rotating black holes must certainly
have $a<m$,  (albeit they can become almost extreme due to accretion processes).
Initially we discuss how to obtain the full
solution of the system of equation~\eref{eq:System of equations for Gordon form}
in the case of the Kerr metric approximated at order $a$; what we obtain is the most general first-order (in $a$) approximation
to the Gordon form of the Kerr metric.
Then we make a consistency check with the Gordon form of the Lense--Thirring metric as expressed in equation~\eref{ansatz Gordon Lense--Thirring}.

\subsubsection{General case of  slow-rotating Kerr in Gordon form
\label{subsec:General-case-of-Gordon-form-in-slowly}} \ \\[5pt]
To obtain the most general Gordon form of the Kerr metric at order
$a$,  one should proceed as described previously in section~\ref{subsec:How-to-find},
transforming the first-order approximation of the Kerr metric, (this is simply equations \eref{eq:Kerr-Schild Kerr}--\eref{eq:Kerr-Schild Kerr background}--\eref{eq:Kerr-Schild Kerr 1-form} with the 1-form $v$ approximated to first order in $a$)
\begin{equation}
\overline{g}_{\rm spheroidal}+\left(-\sqrt{\frac{2m}{r}}\mathrm{d}t\mp\sqrt{\frac{2m}{r}}\mathrm{d}r\mp a\sqrt{\frac{2m}{r}}\sin^{2}\theta\mathrm{d}\phi\right)^{2}+O \left( a^2 \right) \, , \label{KerrMetric order a}
\end{equation}
with the most general coordinate transformation which preserves the explicit
axial symmetry and time translation symmetry, 
\begin{eqnarray}
t & \rightarrow & \sqrt{1-\zeta}\,t
+\tilde{f} \left( r\right)
+a\,G^{t}\left(r,\theta\right)+O\left(a^{2}\right)
\label{eq:General diffeo small a-1}\, ,\\
r & \rightarrow & r+a\,G^{r}\left(r,\theta\right)+O\left(a^{2}\right)
\label{eq:General diffeo small a-2}\, ,\\
\theta & \rightarrow & \theta+\frac{a}{2m}\,G^{\theta}\left(r,\theta\right)+O\left(a^{2}\right)
\label{eq:General diffeo small a-3}\, ,\\
\phi & \rightarrow & \phi+\frac{a}{2m}\,G^{\phi}\left(r,\theta\right)+O\left(a^{2}\right)
\label{eq:General diffeo small a-4}\, .
\end{eqnarray}
Here the function $\tilde{f}$  in equation~\eref{eq:General diffeo small a-1} must (see equation \eref{eq:Gordon Schwarzschild coordinate transformation}) have derivative
\begin{equation}
\tilde{f}^\prime \left(r \right) =\pm \frac{2m}{r-2m} \left( 1- \sqrt{1+\zeta\frac{r-2m}{2m}} \right) \, ,
\end{equation}
for consistency with what we already know is needed, in the case of vanishing $a$, to put the Kerr--Schild form of the Schwarzschild geometry into Gordon form. 
 
In order to find the most general Gordon form of the Kerr metric at
order $a$, the functions $G^{\mu}$ generating the coordinate transformation
must be solutions of the system of equations~\eref{eq:System of equations for Gordon form}
--- after linearization with respect to $a$ --- together with an appropriate
$1$-form $v$.
The solution at zeroth order $a^{0}$ is known, since it will simply be the
Gordon form of the Schwarzschild metric.
At next higher order the $1$-form will include a correction of order $a$ for every component. So we will have
\begin{eqnarray}
\fl
\qquad \qquad v & =-\sqrt{\zeta+\frac{2m}{r}\left(1-\zeta\right)}\mathrm{d}t
\mp\sqrt{\frac{2m}{r}\left(1-\zeta\right)}\mathrm{d}r
+a\,\delta v+O\left(a^{2}\right)\, .
\end{eqnarray}

Taking the first order approximation, the nonlinear system of equations~\eref{eq:System of equations for Gordon form}
is reduced to a system of first-order partial differential equations. This system can be solved
patiently, step by step --- first obtaining the expressions for the components  
$\delta v_{\mu}$ in terms of the functions $G^{\mu}$ and their derivatives,
and then solving the system, finding their explicit expressions. The integration constants should be fixed in such a way that for vanishing
$\zeta$ the $1$-form obtained ultimately reduces to that defining the
Kerr--Schild metric~\eref{KerrMetric order a}.

Here is the full expression obtained for the Gordon form
of the Kerr spacetime, written in the form $g =  \overline{g}_{\rm spheroidal}+v\otimes v$, at first order in $a$:
\begin{eqnarray}
\fl v_{t} & =-\sqrt{\frac{2m}{r}\left(1+\zeta\frac{r-2m}{2m}\right)} \nonumber\\
\fl & \phantom{=}  -\frac{a\kappa}{2r}\sqrt{\frac{2m}{r}\frac{\left(1-\zeta\right)^{2}}{1+\zeta\frac{r-2m}{2m}}}\left(1-\sqrt{1+\zeta\frac{r-2m}{2m}}\right)\cos\theta+O\left(a^{2}\right)\, , \label{small-a vt}\\
\fl v_{r} & = \mp\sqrt{\frac{2m}{r}\left(1-\zeta\right)}\nonumber \\
\fl &\phantom{=} \mp\frac{a\kappa}{2r}\sqrt{\frac{2m}{r}\left(1-\zeta\right)}\left(\left(1-\sqrt{1+\zeta\frac{r-2m}{2m}}\right)+\frac{\zeta r^{2}}{4m^{2}\sqrt{1+\zeta\frac{r-2m}{2m}}}\right)\cos\theta \nonumber \\
\fl & \phantom{=} +O\left(a^{2}\right) \, , \\
\fl v_{\theta} & = \mp a\kappa\sqrt{\frac{2m}{r}\left(1-\zeta\right)}\left(\left(1-\sqrt{1+\zeta\frac{r-2m}{2m}}\right)-\frac{\zeta r\left(r-2m\right)}{8m^{2}\sqrt{1+\zeta\frac{r-2m}{2m}}}\right)\sin\theta   \nonumber \\
\fl & \phantom{=} +O\left(a^{2}\right)\, , \\
\fl v_{\phi} & = \mp a\sqrt{\frac{2m}{r}\frac{1-\zeta}{1+\zeta\frac{r-2m}{2m}}}\sin^{2}\theta +O\left(a^{2}\right)\, .\label{small-a vphi}
\end{eqnarray}

\medskip
\noindent
Here $\kappa$ is a dimensionless residual integration constant one finds from the
coordinate transformation described by equations~\eref{eq:General diffeo small a-1}--\eref{eq:General diffeo small a-4}, 
when the integration constants are chosen to be independent both from $a$ and $\zeta$.
This Gordon form correctly describes the Kerr spacetime up to order $a^2$, \emph{i.e.} it can be  
verified that this Gordon form produces a vanishing Ricci tensor up to  $O(a^2)$. This was expected since we simply considered coordinate
transformations of the Kerr metric; with this check the formalism used is therefore proven to be consistent. 

We observe that in general the norm of the 1-form $v$ is non trivial: for non-vanishing $\kappa$, 
it has a contribution of order $a$ which strongly depends on the angular and radial position
\begin{eqnarray}
\overline{g}^{\mu\nu}v_{\mu}v_{\nu} & =  -\zeta\left(1+\frac{a\kappa}{2m}\frac{1-\zeta}{\sqrt{1+\zeta\frac{r-2m}{2m}}}\cos\theta\right)+O\left(a^{2}\right)\, .
\end{eqnarray}
In the limit of null deformation parameter $\zeta\rightarrow0$, in which case the 1-form $v$
 reproduces the Kerr--Schild case --- the norm vanishes identically, for any value of $\kappa$.

That is, the first order in $a$ Gordon form of the Kerr metric has been obtained as a perturbation of the Gordon form of the Schwarzschild metric,
(which is the limiting case for vanishing $a$). It should be noted that in this more general case the deformation parameter
$\zeta$ is again bounded within the same interval, $\zeta\in\left[0,1\right)$.
If the integration constant $\kappa$ is not neglected the norm is highly
point-dependent; in particular it may be possible that the norm changes
sign due to the presence of the factor $\cos\theta$ in the
term of order $a$. To avoid this change of sign, a bound on $\kappa$
should be imposed. To define this bound we make the assumptions
$0\leq\zeta<1$ and $\left|\frac{a}{M}\right|<1$, and we consider
as the region of interest that with $r>2m$ (implying $1<(1+\zeta\frac{r-2m}{2m})^{1/2}<\infty$).
With such assumptions we find that $\left|\kappa\right|<2$ always prevents
a change of sign of the norm of the $1$-form. In general, for a given
$a/m$ ratio, one needs $\left|\kappa\right|<\left|{2m}/{a}\right|$.

\subsubsection{Consistency check with Lense--Thirring
\label{Consistency check with Lense--Thirring}} \ \\[5pt]
We have found a general expression for the Gordon form of the slow rotating Kerr spacetime. 
We can now easily make a consistency check to prove that this Gordon form, 
defined in terms of the spheroidal flat background and the 1-form of components~\eref{small-a vt}--\eref{small-a vphi}, is equivalent to the Lense--Thirring Gordon form introduced in equations~\eref{eq:Gordon Kerr small a 1-form}~and~\eref{ansatz Gordon Lense--Thirring}.

This can be done assuming a vanishing integration constant $\kappa$ and making use of the properly redefined parameters --- as discussed previously --- of the speed of light in the medium and rescaled mass and angular momentum parameters
\begin{eqnarray}
 \kappa & = 0\, ,\\
 c_*^2 & = 1-\zeta +O\left(a^{2}\right)\, ,\\
\tilde m&= m \frac{1-\zeta}{\zeta}\, ,\\
\tilde a&= \frac{a}{\sqrt{1-\zeta}}\, .
\end{eqnarray}
Extracting the normalization of the 1-form $v$ (for vanishing $\kappa$ its norm is simply $\sqrt{\zeta}$) and
substituting the parameters, we obtain the same expression for $v$ as equation~\eref{eq:Gordon Kerr small a 1-form}
\begin{eqnarray}
\fl \qquad 
V & = \frac{v}{\sqrt{\zeta}}=-\sqrt{1+\frac{2\tilde m}{r}}\mathrm{d}t
\mp\sqrt{\frac{2\tilde m}{r}}\mathrm{d}r
\mp  {2\tilde m\tilde{a}\sin^2\theta\over r\sqrt{1+\frac{2\tilde m}{r}}} \mathrm{d}\phi
+O(\tilde a^{2}) \, .
\end{eqnarray}
Actually, this is not yet quite enough to verify the equivalence between the two Gordon forms because this 1-form is referred to the
spheroidal flat background, while the Lense--Thirring Gordon form in equation~\eref{ansatz Gordon Lense--Thirring} is referred to
the spherical flat metric.
However with a final simple coordinate transformation it is possible to prove that these Gordon forms are completely equivalent:
 The inverse of the transformation from the spherical flat metric to the
 spheroidal flat metric --- approximated at order $a$, and therefore transforming only the coordinate $\phi$ with 
 the inverse of the transformation~\eref{eq: phi transformation} --- is indeed what is needed, since it properly transforms  the
 flat background and does not modify the 1-form up to order $O(a^2)$. This transformation is
 \begin{eqnarray}
\quad \quad \mathrm{d}\phi & \rightarrow\mathrm{d}\phi-\frac{a}{r^2}\mathrm{d}r+O\left(a^2\right) \, ,\\
\overline{g}_{\rm spheroidal} & \rightarrow\overline{g}_{\rm spherical}+O\left(a^2\right) \, ,
\end{eqnarray}
while $V$ (and so $v$) is the same. We have therefore proved the equivalence between these two Gordon forms, they can be obtained one from the
other through a coordinate transformation.

\subsection{Near-null Gordon form of Kerr spacetime}
%

In this section we consider a different approach to the problem of
the description of the Kerr metric in a Gordon form.
We have already seen that --- given a fixed flat background
--- inequivalent Gordon forms of the metric can be obtained
through a class of coordinate transformations which include a rescaling
of the time coordinate. To describe such deformations we have used
the rescaling term $\sqrt{1-\zeta}$.
Rescaling the time coordinate, one can obtain new $1$-forms
with non-null norm, passing from a  Kerr--Schild form
to a Gordon form. The parameter $\zeta$ is therefore
strictly related to the norm and becomes the instrument to explore
the space of possible inequivalent Gordon forms of the Kerr
spacetime.

In the slow-rotation case, the order $a$ approximation has allowed
us to consider Gordon forms where the rescaling parameter $\zeta$
was free to explore the whole range of allowed values: That is, the same interval
$\zeta\in\left[0,1\right)$ which was acceptable in the Schwarzschild case. While the small $a$ 
result is already of great interest, it rules out the whole regime of rapidly
rotating spacetimes.
We now wish to explore that region of parameter space, and we want
to do so by making a different approximation: We consider a small deformation
parameter $\zeta$, and we obtain an expression for the Gordon form of the Kerr
metric at first order in $\zeta$.

\subsubsection{Infinitesimal local translation of the Kerr metric}\ \\[5pt]
 \label{sec:Lie derivative method}
We want to apply the procedure described in the introductory subsection~\eref{subsec:How-to-find}, transforming the Kerr--Schild
form of the Kerr metric~\eref{eq:Kerr-Schild Kerr} with an infinitesimal
transformation.
Again we want this coordinate transformation to include a deformation of the
time coordinate; since we want this deformation to be infinitesimal,
we can approximate at order $\zeta$ the rescaling term, and consider
$\zeta$ to be arbitrary small 
\begin{eqnarray}
\sqrt{1-\zeta} & =1-\frac{\zeta}{2}+O\left(\zeta^{2}\right).
\end{eqnarray}
The other coordinates should be transformed infinitesimally too, and
the rest of the transformation should be assumed not to spoil the
explicit axial and time translation symmetries. So we consider
\begin{eqnarray}
t & \rightarrow & \left(1-\frac{\zeta}{2}\right)t+\zeta\, F^{t}\left(r,\theta\right)+O\left(\zeta^{2}\right)\label{eq:Gordon Kerr near-null coordinate transformation-1}\, ,\\
r & \rightarrow & r+\zeta\, F^{r}\left(r,\theta\right)+O\left(\zeta^{2}\right)
\label{eq:Gordon Kerr near-null coordinate transformation-2}\, ,\\
\theta & \rightarrow & \theta+\zeta\, F^{\theta}\left(r,\theta\right)+O\left(\zeta^{2}\right)
\label{eq:Gordon Kerr near-null coordinate transformation-3}\, ,\\
\phi & \rightarrow & \phi+\zeta\, F^{\phi}\left(r,\theta\right)+O\left(\zeta^{2}\right)
\label{eq:Gordon Kerr near-null coordinate transformation-4}\, .
\end{eqnarray}
This coordinate transformation is easily brought back to the formalism of infinitesimal
local translations as presented in section~\eref{subsec:How-to-find}. The coordinates
are transformed with $x^{\mu}\rightarrow x^{\mu}+\xi^{\mu}\left(x\right)$,
and the translation vector $\xi$ defining this coordinate transformation  is
clearly of order $\zeta$. This means that for an arbitrarily
small $\zeta$, the Kerr metric~\eref{eq:Kerr-Schild Kerr} is moved
infinitesimally along the vector field $\xi$, and for this infinitesimal
transformation it can be written in terms of the Lie derivative along this vector field
\begin{eqnarray}
g_{\mu\nu} & \rightarrow & g_{\mu\nu}+\mathcal{L}_{\xi}g_{\mu\nu}=g_{\mu\nu}+\left(\mathcal{L}_{\xi}\overline{g}_{\mu\nu}+v_{\mu}\mathcal{L}_{\xi}v_{\nu}+\mathcal{L}_{\xi}v_{\mu}v_{\nu}\right)\label{eq:Kerr metric pushforward}\, ,
\end{eqnarray}
where we are now dropping the subscript in the flat background metric,
writing $\overline{g}$ instead of $\overline{g}_{\rm spheroidal}$ for simplicity.
Now of course the relevant question is: Can the transformed expression of the Kerr metric be written in Gordon
form? This will be the focus of the next section.

\subsubsection{Solution in modified components}\ \\[5pt]
The infinitesimal translation transforms the Kerr--Schild
form metric~\eref{eq:Kerr-Schild Kerr} into the modified metric tensor~\eref{eq:Kerr metric pushforward} through the Lie derivative; both
are metric tensors describing the Kerr spacetime with the same set
of coordinates. The new expression can be put in a Gordon form if
it is possible to find a new $1$-form which solves, together with
vector field $\xi$, the system of equations~\eref{eq:System of equations for Gordon form}.

The new $1$-form must differ from that defining the initial Kerr--Schild
form in equation~\eref{eq:Kerr-Schild Kerr 1-form} by a term of order $\zeta$,
such that the system admits a solution valid for every small value
of the deformation parameter; \emph{i.e.} if the Kerr--Schild form was
given (with respect to the spheroidal flat background) by the 1-form
\begin{eqnarray}
v & =\Phi \left(- \mathrm{d}t \mp \mathrm{d}r \mp  a\sin^{2}\theta \mathrm{d}\phi \right) \, ,\\
\Phi & =\sqrt{\frac{2mr}{r^{2}+a^{2}\cos^{2}\theta}}\, ,
\end{eqnarray}
where the function $\Phi$ is a shorthand defined for convenience. Then the 
new solution must be of the form $v+\delta v$, with $\delta v$
being the correction of order $\zeta$.
With this assumption for the modified metric tensor and the $1$-form,
 equation~\eref{eq:System of equations for Gordon form} is expressed
at order $\zeta$ as
\begin{eqnarray}
\fl
\qquad g_{\mu\nu}+\left(\mathcal{L}_{\xi}\overline{g}_{\mu\nu}
+v_{\mu}\;\mathcal{L}_{\xi}v_{\nu}+\mathcal{L}_{\xi}v_{\mu}\; v_{\nu}\right)-\overline{g}_{\mu\nu} 
& 
=  \left(v_{\mu}+\delta v_{\mu}\right)\left(v_{\nu}+\delta v_{\nu}\right)\, ,\\
& \Downarrow\nonumber \\
 \fl
\qquad\qquad\qquad\qquad
\mathcal{L}_{\xi}\overline{g}_{\mu\nu}
+v_{\mu}\;\mathcal{L}_{\xi}v_{\nu}+\mathcal{L}_{\xi}v_{\mu}\; v_{\nu} & 
=  v_{\mu}\,\delta v_{\nu}+\delta v_{\mu}\, v_{\nu}+O\left(\zeta^{2}\right)\, .
\label{eq:reduced}
\end{eqnarray}
It is easier to solve this problem by change of dependent variables,
redefining the $1$-form of interest. If we consider the modified 1-form correction
\begin{eqnarray}
\delta v^{\prime} & =\delta v-\mathcal{L}_{\xi}v,
\end{eqnarray}
then equation~\eref{eq:System of equations for Gordon form}, which we have already reduced to \eref{eq:reduced}, 
takes the simplified form
\begin{eqnarray}
\mathcal{L}_{\xi}\overline{g}_{\mu\nu} & 
=v_{\mu}\, \delta v_{\nu}^{\prime}+\delta v_{\mu}^{\prime}\,v_{\nu}+O\left(\zeta^{2}\right).
\label{eq:Simplified Gordon problem for Kerr}
\end{eqnarray}
For this system, similarly to what has been done for the general
case of the Gordon form in the slowly rotating approximation for Kerr spacetime as presented in 
section~\eref{subsec:General-case-of-Gordon-form-in-slowly}, the solution
is found step by step. 
First we obtain the expressions for the
components of $\delta v^{\prime}$ in terms of the components 
--- and their derivatives --- of the translation
vector field $\xi$.  Then the system is solved by sequentially finding the functions $\xi^{\mu}$ one
after the other. The integration constants should be chosen such that
the coordinate transformation reduces to the identity for vanishing $\zeta$,
ensuring that no trivial translations in the coordinates $t$ and
$\phi$ are introduced.
Doing all this, the components of the translation vector field are found (to order $O(\zeta)$) to be
\begin{eqnarray}
\xi^{t} & =  \left(-\frac{1}{2}t \mp \frac{r^{2}+a^{2}\cos^{2}\theta}{2r}\right)\zeta\, ,\\
\xi^{r} & =  \left(\lambda \frac{\left(r^{2}+a^{2}\right)\cos\theta}{r^{2}+a^{2}\cos^{2}\theta}+\frac{a^{2}}{2}\left(\frac{\cos^{2}\theta}{r}-\frac{r\sin^{2}\theta}{r^{2}+a^{2}\cos^{2}\theta}\right)\right)\zeta \, ,\\
\xi^{\theta} & =  \left(- \lambda \frac{r\sin\theta}{r^{2}+a^{2}\cos^{2}\theta}-\frac{a^{2}}{2}\frac{\sin\theta\cos\theta}{r^{2}+a^{2}\cos^{2}\theta}\right)\zeta \, ,\\
\xi^{\phi} & =  \left(-\lambda \frac{a\cos\theta}{r^{2}+a^{2}\cos^{2}\theta}+\frac{a}{2}\frac{r}{r^{2}+a^{2}\cos^{2}\theta}\right)\zeta \,.
\end{eqnarray}
Here $\lambda$ is a dimensionful integration constant.
Inserting these back into the expressions for the $\delta v^{\prime}$, and evaluating the Lie derivative of the 1-form $v$,
we obtain the solutions (to order $O(\zeta)$) for the components of the modified $1$-form $v+\delta v$ on the spehoridal flat background $\overline{g}=\overline{g}_{\rm spheroidal}$.
We first note
\begin{eqnarray}
v_{\mu}+\delta v_{\mu} & =v_{\mu}+\delta v_{\mu}^{\prime}+\mathcal{L}_{\xi}v_{\mu}=
 v_{\mu}+\delta v_{\mu}^{\prime}+\xi^{\sigma}\partial_{\sigma}v_{\mu}+v_{\sigma}\partial_{\mu}\xi^{\sigma}\, .
\end{eqnarray}
The components of $\delta v^{\prime}$ are easily found by substitution after solving equation~\eref{eq:Simplified Gordon problem for Kerr}. We find
\begin{eqnarray}
\delta v^{\prime} & =\frac{\zeta}{2\Phi}\left(-\mathrm{d}t
\pm\frac{a^{2}\cos^{2}\theta}{r^{2}}\mathrm{d}r
\pm\frac{a^{2}\sin2\theta}{r}\mathrm{d}\theta
\pm a\sin^{2}\theta\mathrm{d}\phi\right)+O\left(\zeta^{2}\right) \, .
\end{eqnarray}
The other contribution to $\delta v$ comes from the Lie derivatives of the 1-form $v$ with respect to the translation vector field $\xi$, and in general depends on the integration constant $\lambda$
\begin{eqnarray}
\mathcal{L}_{\xi}v=&\left(\mathcal{L}_{\xi}\Phi\right)\left(-\mathrm{d}t\mp\mathrm{d}r\mp a\sin^{2}\theta\mathrm{d}\phi\right) \nonumber \\
& +\frac{\Phi}{2}\left(\mathrm{d}t\pm\mathrm{d}r\pm2\frac{a^{2}r\cos\theta\sin\theta}{r^{2}+a^{2}\cos^{2}\theta}\mathrm{d}\theta\pm2\frac{a^{3}\cos^{2}\theta\sin^{2}\theta}{r^{2}+a^{2}\cos^{2}\theta}\mathrm{d}\phi\right) \nonumber \\
& \pm\lambda\Phi\left(\frac{r^{2}-a^{2}\cos^{2}\theta}{r^{2}+a^{2}\cos^{2}\theta}\mathrm{d}\theta+2\frac{ar\cos\theta\sin\theta}{r^{2}+a^{2}\cos^{2}\theta}\mathrm{d}\phi\right) \, , \\
\end{eqnarray}
where
\begin{eqnarray}
\mathcal{L}_{\xi}\Phi=&-\Phi\frac{\left(2r^{4}a^{2}\cos\left(2\theta\right)+3r^{2}a^{4}\sin\left(2\theta\right)-2a^{6}\cos^{6}\theta\right)}{8r^{2}\left(a^{2}\cos^{2}\theta+r^{2}\right)^{2}} \nonumber \\
& -\lambda\Phi\frac{\left(r^{4}+3r^{2}a^{2}\sin^{2}\theta-a^{4}\cos^{2}\theta\right)}{2r\left(a^{2}\cos^{2}\theta+r^{2}\right)^{2}}\cos\theta \, .
\end{eqnarray}

It is interesting to note that the components $\delta v^{\prime}$
can be interpreted as providing the components of the same $1$-form
as $\delta v$, when the chosen background is not $\overline{g}$
but $\overline{g}-\mathcal{L}_{\xi}\overline{g}$. This tensor, differing
from the flat background by its Lie derivative is still an acceptable
approximately flat background (up to order $\zeta$ in the metric and order $\zeta^2$ in the Riemann tensor) in order to put the Kerr metric in a Gordon form.
Manipulating the expression of the metric~\eref{eq:Kerr metric pushforward}
with the solution found from equation~\eref{eq:Simplified Gordon problem for Kerr},
we can verify that
\begin{eqnarray}
g_{\mu\nu} & =\left(\overline{g}_{\mu\nu}-\mathcal{L}_{\xi}\overline{g}_{\mu\nu}\right)+\left(v_{\mu}+\delta v_{\mu}^{\prime}\right)\left(v_{\nu}+\delta v_{\nu}^{\prime}\right)\, .
\end{eqnarray}
This means that the Kerr metric in Kerr--Schild form can be
manipulated with Lie derivatives with respect to the vector
field $\xi$ we obtained, in order to write it down as a Gordon form with a modified
flat background and a modified $1$-form. It is in this setup that
is more convenient to evaluate the norm of the $1$-form, which can
be evaluated straightforwardly by the known inverse of the metric
\begin{eqnarray}
g^{\mu\nu}\left(v_{\mu}+\delta v_{\mu}^{\prime}\right)\left(v_{\nu}+\delta v_{\nu}^{\prime}\right) & 
=-\zeta\;\frac{r^{2}+a^{2}\cos^{2}\theta}{r^{2}}+O\left(\zeta^{2}\right)\, ,
\end{eqnarray}
proving that this is really a Gordon form for a near null $1$-form.

\section{Vorticity and applications to analogue spacetimes}
Interest in the Gordon form of spacetime metrics is due (among other things) to 
potential applications in the analogue spacetime programme. See specifically~\cite{Visser:2010,Fagnocchi:2010sn}, and more generally~\cite{LRR, Visser:2001,Visser:2007,Visser:2013,Visser:2004,Visser:2005,Finke:2016,Braden:2017}. Indeed the Gordon form, (or something conformal to the Gordon form),  generically describes the acoustic metric experienced by a linearised perturbation on a relativistic fluid~\cite{Visser:2010,Fagnocchi:2010sn}.
That one might want vorticity in analogue systems is clear from references~\cite{Visser:2004b,Richartz:2014,Cardoso:2016,Torres:2016,Torres:2017,Patrick:2018,Cropp:2015,Giacomelli:2017}. 
Very often, however, in theoretical analyses of these analogue systems the inclusion of vorticity is tricky~\cite{Perez-Bergliaffa:2001} --- most typically the four velocity of the fluid is by construction hypersurface orthogonal (implying that it can be written as being proportional to the gradient of some scalar function) and as such --- by the Frobenius theorem --- it is vorticity free (in the relativistic sense that $V\wedge \d V=0$). This is potentially a problem for an experimental simulation of a true Kerr geometry given that one can very easily realise that the Kerr--Schild and Kerr--Gordon forms of the Kerr metric found in this paper always require a four velocity (or equivalently a one-form) which is not vorticity free.
This fact can be seen most easily by looking at the simplest case, the four velocity obtained in the Gordon form of the Lense--Thirring spacetime,
\begin{equation}
V =-\sqrt{1+\frac{2\tilde{m}}{r}}\mathrm{d}t\mp\sqrt{\frac{2\tilde{m}}{r}}\mathrm{d}r\mp\tilde{a}\sqrt{\frac{2\tilde{m}}{r}\frac{2\tilde{m}}{r+2\tilde{m}}}\sin^{2}\theta\mathrm{d}\phi+O\left(a^{2}\right).
\end{equation}
For this four velocity we can compute the 4-vorticity
\begin{eqnarray}
\fl \omega = \star\left(V\wedge\mathrm{d}V\right)	
& =V_{\nu_{1}}V_{\nu_{3},\nu_{2}}\; \star\left(\mathrm{d}x^{\nu_{1}}\wedge\mathrm{d}x^{\nu_{2}}\wedge\mathrm{d}x^{\nu_{3}}\right)\\
\fl 	& =V_{\nu_{1}}V_{\nu_{3},\nu_{2}}\; \epsilon^{\nu_{1}\nu_{2}\nu_{3}\nu_{4}}\overline{g}_{\nu_{4}\mu_{4}}\;\mathrm{d}x^{\mu_{4}}\\
\fl 	& =\sqrt{\overline{g}}V_{\nu_{1}}V_{\nu_{3},\nu_{2}}\;\overline{g}^{\nu_{1}\mu_{1}}\overline{g}^{\nu_{2}\mu_{2}}\overline{g}^{\nu_{3}\mu_{3}}\; \hat{\epsilon}_{\mu_{1}\mu_{2}\mu_{3}\mu_{4}}\;\mathrm{d}x^{\mu_{4}}\\
\fl 	& =-\frac{4\tilde{a}\tilde{m}\cos\theta}{r^{3}}\sqrt{\frac{2\tilde{m}}{r+2\tilde{m}}}\mathrm{d}t\mp\frac{4\tilde{a}\tilde{m}\cos\theta}{r^{3}}\mathrm{d}r\mp\frac{2\tilde{a}\tilde{m}\sin\theta}{r\left(2\tilde{m}+r\right)}\mathrm{d}\theta\, .\label{eq:vorticity}
\end{eqnarray}
The 4-vorticity 1-form $\omega$ is evaluated with respect to the usual flat background.
This result is non vanishing at first order in $a$, and is valid for arbitrary deformation parameter $\zeta\in(0,1)$. This is enough to imply that also for the near-null Gordon form 
it is impossible for the 4-vorticity to vanish. Indeed, the expression of the 4-vorticity of the near null Gordon form will depend on the parameter $a$; at first order, it must be consistent with the non-vanishing expression in equation~\eref{eq:vorticity}. It will therefore also be impossible for the 4-vorticity in the near-null case to vanish identically.

Hence we conclude that any analogue model of the Kerr geometry will have to necessarily include vorticity in the background flow. Models of this sort are indeed available and are discussed in a companion paper~\cite{Vorticity}.

%
\section{Further developments --- towards a full solution}
We have presented two approximate results for the Gordon form of the Kerr metric, describing 
the regimes of small rotation and near-null 1-form; they were obtained respectively by perturbing
the Gordon form of the Schwarzschild metric, and by infinitesimally deforming the Kerr--Schild form of the Kerr metric.
These two results plausibly suggest the existence of an (as yet unknown) full analytical expression for the Gordon form of the Kerr metric, one which should reproduce 
the results presented herein when considering the first order approximations in the two parameters $a$ and $\zeta$.
Such an expression would be the full solution of the system of equations~\eref{eq:System of equations for Gordon form}, possibly found  
through the algorithm presented in subsection \ref{subsec:How-to-find}.

Finally, we remark that in finding these results we found that a proper choice of flat background is of 
crucial importance for the resolution of the problem  --- both in terms of the final expression for the Gordon form, and in computation time,
and this should be considered in any future approach to this problem.
We hope to further explore these and other possibilities in the next future.

\section*{Acknowledgments}
MV was supported by the Marsden Fund, 
which is administered by the Royal Society of New Zealand.
MV would also like to thank both SISSA and INFN (Trieste) for hospitality during the initial phase of this work.
\section*{References}
  

\begin{thebibliography}{69}  

  
\bibitem{Gordon}
Walter Gordon, ``Zur Lichtfortpflanzung nach der Relativit\"atstheorie'', \\
Annalen der Physik {\bf 377(22)} (1923) 421--456; doi: 10.1002/andp.19233772202\\
``On the propagation of light in the theory of relativity'',
translated by D. H. Delphenich. \\
{\sf http://www.neo-classical-physics.info/uploads/3/4/3/6/34363841/gordon\_-\_optical\_metrics.pdf}

\bibitem{Landau-Lifshitz}
Landau, L.D. and Lifshitz, E.M., \emph{The classical theory of fields}, (Pergamon Press, Oxford, 1971), 3rd edition.


\bibitem{Plebanski:1960}
J.~Pleba\'nski, ``Electromagnetic waves in gravitational fields'', Phys. Rev., 118, 1396--1408, (1960). 

\bibitem{Plebanski:1970}
J.~Pleba\'nski, \emph{Lectures on Nonlinear Electrodynamics}, (Nordita, Copenhagen, 1970).
  
\bibitem{deFelice:1971}
F.~de Felice, ``On the gravitational field acting as an optical medium'', \\
Gen. Relativ. Gravit., 2, 347--357, (1971).

\bibitem{Skrotski}
Skrotskii, G.V., ``The influence of gravitation on the propagation of light'', \\
Sov. Phys. Dokl., 2, 226--229, (1957).

\bibitem{Balzas}
Balazs, N.L., ``Effect of a gravitational field, due to a rotating body, on the plane of polarization of an electromagnetic wave'', Phys. Rev., 110, 236--239, (1958).

\bibitem{Anderson}
Anderson, J.L. and Spiegel, E.A., ``Radiative transfer through a flowing refractive medium'', Astrophys. J., 202, 454--464, (1975). 


\bibitem{Pham}
Pham, Q.M., ``Sur les \'equations de l'electromagn\'e dans la materie'', \\
C. R. Hebd. Seanc. Acad. Sci., 242, 465--467, (1956).



\bibitem{Schuster:2017}
  S.~Schuster and M.~Visser,\\
  ``Effective metrics and a fully covariant description of constitutive tensors in electrodynamics'',\\
Physical Review D {\bf96} (2017)124019 
doi: 10.1103/PhysRevD.96.124019\\{}  [arXiv:1706.06280 [gr-qc]].
  
  \bibitem{Schuster:2018}
  S.~Schuster and M.~Visser,\\
  ``Bespoke analogue space-times: Meta-material mimics'',
  arXiv:1801.05549 [gr-qc].
  
\bibitem{Vorticity}
  S.~Liberati, S.~Schuster, G.~Tricella, and M.~Visser,
  ``Vorticity in analogue spacetimes'',\\
  arXiv:1802.04785 [gr-qc].
  
 \bibitem{Boyer-Lindquist}
 S.~Schuster and M.~Visser,
 ``Boyer-Lindquist space-times and beyond: Meta-material analogues,''
 arXiv:1802.09807 [gr-qc].


\bibitem{Rosquist}
K.~Rosquist,
  ``A Moving medium simulation of Schwarzschild black hole optics'',\\
  Gen.\ Rel.\ Grav.\  {\bf 36} (2004) 1977
  doi: 10.1023/B:GERG.0000036055.82140.06
  [gr-qc/0309104].
  

\bibitem{Liberati}
L.~Giacomelli and S.~Liberati,
  ``Rotating black hole solutions in relativistic analogue gravity'',\\
  Phys.\ Rev.\ D {\bf 96} (2017) no.6,  064014
  doi:10.1103/PhysRevD.96.064014\\{}
  [arXiv:1705.05696 [gr-qc]].
  


\bibitem{Kerr}
R.~P.~Kerr,
  ``Gravitational field of a spinning mass as an example of algebraically special metrics'',
  Phys.\ Rev.\ Lett.\  {\bf 11} (1963) 237.
  doi:10.1103/PhysRevLett.11.237

  
\bibitem{Kerr-book}
D.~L.~Wiltshire, M.~Visser and S.~M.~Scott (editors),
  ``The Kerr spacetime: Rotating black holes in general relativity'',
  (Cambridge University Press, Cambridge, England, 2009)

\bibitem{Kerr-intro}
M.~Visser,
  ``The Kerr spacetime: A brief introduction'',
  arXiv:0706.0622 [gr-qc].
  Published in~\cite{Kerr-book}.
  
  \bibitem{Teukolsky:2014}
  S.~A.~Teukolsky,
  ``The Kerr metric'',
  Class.\ Quant.\ Grav.\  {\bf 32} (2015) no.12,  124006
  doi:10.1088/0264-9381/32/12/124006
  [arXiv:1410.2130 [gr-qc]].

  \bibitem{Doran}
  C.~Doran,
  ``A new form of the Kerr solution'',
  Phys.\ Rev.\ D {\bf 61} (2000) 067503\\
  doi:10.1103/PhysRevD.61.067503
  [gr-qc/9910099].
  
  \bibitem{River}
  A.~J.~S.~Hamilton and J.~P.~Lisle,
  ``The river model of black holes'',\\
  Am.\ J.\ Phys.\  {\bf 76} (2008) 519
  doi:10.1119/1.2830526
  [gr-qc/0411060].
  
  
\bibitem{Visser:2010}
  M.~Visser and C.~Molina-Par\'is,\\
  ``Acoustic geometry for general relativistic barotropic irrotational fluid flow'',\\
  New J.\ Phys.\  {\bf 12} (2010) 095014
  doi:10.1088/1367-2630/12/9/095014\\{}
  [arXiv:1001.1310 [gr-qc]].
  

\bibitem{Unruh:1980}
  W.~G.~Unruh,
  ``Experimental black hole evaporation'',
  Phys.\ Rev.\ Lett.\  {\bf 46} (1981) 1351.\\
  doi:10.1103/PhysRevLett.46.1351

 \bibitem{Visser:1993}
  M.~Visser,
  ``Acoustic propagation in fluids: An unexpected example of Lorentzian geometry'',\\
  gr-qc/9311028.

\bibitem{Visser:1997}
  M.~Visser,
  ``Acoustic black holes: Horizons, ergospheres, and Hawking radiation'',\\
  Class.\ Quant.\ Grav.\  {\bf 15} (1998) 1767
  doi:10.1088/0264-9381/15/6/024
  [gr-qc/9712010].
  
  
   \bibitem{Barcelo:2000}
  C.~Barcel\'o, S.~Liberati and M.~Visser,
  ``Analog gravity from Bose-Einstein condensates'',\\
  Class.\ Quant.\ Grav.\  {\bf 18} (2001) 1137
  doi:10.1088/0264-9381/18/6/312
  [gr-qc/0011026].
  
  \bibitem{Barcelo:2003}
  C.~Barcel\'o, S.~Liberati and M.~Visser,
  ``Probing semiclassical analog gravity in Bose-Einstein condensates with widely tunable interactions'',\\
  Phys.\ Rev.\ A {\bf 68} (2003) 053613
  doi:10.1103/PhysRevA.68.053613
  [cond-mat/0307491].




  


\bibitem{LRR}
 C.~Barcel\'o, S.~Liberati and M.~Visser,
  ``Analogue gravity'',
  Living Rev.\ Rel.\  {\bf 8} (2005) 12\\{}
   [Living Rev.\ Rel.\  {\bf 14} (2011) 3]
  doi: 10.12942/lrr-2005-12
  [gr-qc/0505065].
  
  \bibitem{Visser:2001}
  M.~Visser, C.~Barcel\'o and S.~Liberati,
  ``Analog models of and for gravity'',\\
  Gen.\ Rel.\ Grav.\  {\bf 34} (2002) 1719
  doi:10.1023/A:1020180409214
  [gr-qc/0111111].
  
  \bibitem{Visser:2007}
  M.~Visser and S.~Weinfurtner,
  ``Analogue spacetimes: Toy models for ''quantum gravity'''',
  PoS QG {\bf -PH} (2007) 042
  [arXiv:0712.0427 [gr-qc]].

  
  \bibitem{Visser:2013}
  M.~Visser,
  ``Survey of analogue spacetimes'',\\
  Lect.\ Notes Phys.\  {\bf 870} (2013) 31
  doi:10.1007/978-3-319-00266-8\_2
  [arXiv:1206.2397 [gr-qc]].
  
\bibitem{Fagnocchi:2010sn}
  S.~Fagnocchi, S.~Finazzi, S.~Liberati, M.~Kormos and A.~Trombettoni,\\
  ``Relativistic Bose-Einstein condensates: a new system for analogue models of gravity'',\\
  New J.\ Phys.\  {\bf 12} (2010) 095012
  doi:10.1088/1367-2630/12/9/095012
  [arXiv:1001.1044 [gr-qc]].
  

 \bibitem{Visser:2004}
  M.~Visser and S.~Weinfurtner,
  ``Massive phonon modes from a BEC-based analog model'',\\
  cond-mat/0409639.
  
  \bibitem{Visser:2005}
  M.~Visser and S.~Weinfurtner,\\
  ``Massive Klein-Gordon equation from a BEC-based analogue spacetime'',\\
  Phys.\ Rev.\ D {\bf 72} (2005) 044020
  doi:10.1103/PhysRevD.72.044020
  [gr-qc/0506029].
  


  \bibitem{Finke:2016}
  A.~Finke, P.~Jain and S.~Weinfurtner,\\
  ``On the observation of nonclassical excitations in Bose--Einstein condensates'',\\
  New J.\ Phys.\  {\bf 18} (2016) no.11,  113017
  doi:10.1088/1367-2630/18/11/113017\\{}
  [arXiv:1601.06766 [quant-ph]].

  \bibitem{Braden:2017}
  J.~Braden, M.~C.~Johnson, H.~V.~Peiris and S.~Weinfurtner,\\
  ``Towards the cold atom analog false vacuum'',
  arXiv:1712.02356 [hep-th].

  
  \bibitem{Visser:2004b}
  M.~Visser and S.~Weinfurtner,
  ``Vortex geometry for the equatorial slice of the Kerr black hole'',
  Class.\ Quant.\ Grav.\  {\bf 22} (2005) 2493
  doi:10.1088/0264-9381/22/12/011
  [gr-qc/0409014].
  
  \bibitem{Richartz:2014}
  M.~Richartz, A.~Prain, S.~Liberati and S.~Weinfurtner,\\
  ``Rotating black holes in a draining bathtub: super-radiant scattering of gravity waves'',\\
  Phys.\ Rev.\ D {\bf 91} (2015) no.12,  124018
  doi:10.1103/PhysRevD.91.124018\\{}
  [arXiv:1411.1662 [gr-qc]].
  
  \bibitem{Cardoso:2016}
  V.~Cardoso, A.~Coutant, M.~Richartz and S.~Weinfurtner,\\
  ``Detecting Rotational Super-radiance in Fluid Laboratories'',\\
  Phys.\ Rev.\ Lett.\  {\bf 117} (2016) no.27,  271101\\
  doi:10.1103/PhysRevLett.117.271101
  [arXiv:1607.01378 [gr-qc]].

  
  \bibitem{Torres:2016}
  T.~Torres, S.~Patrick, A.~Coutant, M.~Richartz, E.~W.~Tedford and S.~Weinfurtner,\\
  ``Observation of super-radiance in a vortex flow'',
  Nature Phys.\  {\bf 13} (2017) 833\\
  doi:10.1038/nphys4151
  [arXiv:1612.06180 [gr-qc]].
  
  \bibitem{Torres:2017}
  T.~Torres, A.~Coutant, S.~Dolan and S.~Weinfurtner,\\
  ``Waves on a vortex: rays, rings and resonances'',
  arXiv:1712.04675 [gr-qc].
  
  \bibitem{Patrick:2018}
  S.~Patrick, A.~Coutant, M.~Richartz and S.~Weinfurtner,\\
  ``Black hole quasi-bound states from a draining bathtub vortex flow'',
  arXiv:1801.08473 [gr-qc].
  

   
  \bibitem{Cropp:2015}
  B.~Cropp, S.~Liberati and R.~Turcati,
  ``Vorticity in analog gravity'',\\
  Class.\ Quant.\ Grav.\  {\bf 33} (2016) no.12,  125009\\
  doi:10.1088/0264-9381/33/12/125009
  [arXiv:1512.08198 [gr-qc]].
  
  \bibitem{Giacomelli:2017}
  L.~Giacomelli and S.~Liberati,
  ``Rotating black hole solutions in relativistic analogue gravity'',\\
  Phys.\ Rev.\ D {\bf 96} (2017) no.6,  064014
  doi:10.1103/PhysRevD.96.064014 \\{}
  [arXiv:1705.05696 [gr-qc]].
  
\bibitem{Perez-Bergliaffa:2001}
  S.~E.~Perez Bergliaffa, K.~Hibberd, M.~Stone and M.~Visser,\\
  ``Wave equation for sound in fluids with vorticity'',\\
  Physica D {\bf 191} (2004) 121
  doi:10.1016/j.physd.2003.11.007
  [cond-mat/0106255].
  




\end{thebibliography}
\end{document}